\documentclass[twocolumn,floatfix]{revtex4}
\usepackage{epsf,graphicx}
\usepackage{amsmath,amssymb,amsfonts}
\begin{document}
\title{Flaw of Jarzynski's equality when applied to systems\\ with several degrees of freedom. }
\author{D.H.E. Gross}
\affiliation{Hahn-Meitner Institute and Freie Universit{\"a}t
Berlin, Fachbereich Physik. Glienickerstr. 100; 14109 Berlin,
Germany} \email{ gross@hmi.de}
\homepage{http://www.hmi.de/people/gross/}

\begin{abstract}
Simple example: During the sudden expansion of an isolated ideal gas from a
small volume $V_0$ into a larger one $V_1$, the entropy changes by $\Delta
S= N\ln(V_1/V_0)=-\beta\Delta F>0$ but no work $W$ is produced nor
absorbed. Consequently, Jarzynski's identity $\left<\exp(-\beta W)\right>=
­ \exp(-\beta \Delta F)) $ is wrong.
\end{abstract}
\maketitle

In his paper \cite{jarzynski99} Jarzynski claims the equation (J51)
\begin{equation}
\left<\exp(-\beta W)\right>=  \exp(-\beta \Delta
F)\hspace{2.5cm}(J51)\label{jarzynski}
\end{equation}to be valid independently of whether the path from the initial equilibrium  state $0$ to
the final equilibrium state $1$ of the system is fast or slow. Here $W$ is
the work done on the system $\psi$ in each realization of the process, $<>$
means the average over many realizations in contact with an external heat
bath $\theta$ at temperature $T=1/\beta$ and $\Delta F=F_1-F_0$ is the
difference of the free energies of the system between its final and initial
equilibrium states.
\section{Trivial counterexample} Already Sung gave a trivial
counter example to Jarzynski's identity \cite{sung05}: An ideal
gas of $N$ non-interacting point particles with mass $m$ at an
energy $E$ in the volume $V$ has the (Boltzmann) entropy
\begin{equation}
S_{id-gas}=\ln\left\{\frac{E^{(3N-2)/2}{V^N }(m)^{3N/2}}{N!\Gamma(3N/2)
(2\pi\hbar^2)^{3N/2}}\right\},
\end{equation} consequently, in an iso-energetic sudden expansion it increases by
\begin{equation}
\Delta S=-\beta\Delta F=N\ln(V_1/V_0).
\end{equation}Due to the non-existence of any interaction among the
particles, the energy of the gas does not change and no work is performed
($W\equiv 0$). Also the (microcanonical) temperature of the gas remains
constant. I.e. ``equation''(\ref{jarzynski}) is wrong.

Now, this is a microcanonical system and some claim relation
(\ref{jarzynski}) cannot be applied to microcanonical ensembles. However,
in \cite{jarzynski99} Jarzynski writes in the discussion: {\em ``The
derivation explicitly assumes that the reservoir degrees of freedom are
initially sampled canonically. The result itself, however, might not depend
on this assumption as the derivation suggests. For macroscopic large
reservoir, Eq.(J4),in \cite{jarzynski99}, ought to remain valid at least to
an excellent approximation, if the initial conditions are sampled from
microcanonical, rather than canonical, distribution.''} Our example clearly
shows it {\em does} depend on the assumption of a coupling to a canonical
bath even though our bath does not matter at all as the process does not
produce any heat and is absolutely isoenergetical. Of course, the lack of a
heat bath may not be the case Jarzynski considers in this statement about
the equivalence. However, as said above in this example the heat bath does
not matter at all. Thus I think this is not the reason why his equation
does not work here.

So where is the flaw in Jarzynski's "proof"? Jarzynski considers
``\underline{deterministic}'' paths $\{z(t),Y(t)\}\equiv\Gamma(t)$ in the
\{$6N$+bath-dofs.\}-dimensional phase space $\Psi$ of the system $\psi$
with phase space point $z(t)$, and the bath $\theta$ with phase space point
$Y(t)$ from the initial point $\Gamma_0,$ to the final $\Gamma_1$ in the
finite switching time $\tau$.

Jarzynski writes: {\em ``$\Delta S$ can be obtained by projecting out from
the complete $\Gamma(t)$ the reservoir degrees of freedom (dof); by using
the initial and final conditions $\Gamma(0)$ and $\Gamma(\tau)$ to compute
the net change $\Delta Q$ in the internal energy of the reservoir $\Delta Q
$''}. Then he uses Clausius formula
\begin{equation}
\Delta S=\Delta S_b= \Delta Q/T,\hspace{2.5cm}(J2)\label{clausius}
\end{equation} to relate the energy transfer to the bath to its entropy
generation $\Delta S_b$.

He continues by writing:{\em``The ensemble of realizations which we would
obtain by endlessly repeating the same process, always initializing $\psi$
in the microstate $z_A$; the differences from one realization to the next
arises solely from the different \underline{initial} conditions sampled
from the reservoir.''} From this ensemble he constructs the joint
probability $P(z_B,\Delta S|z_A)$ to obtain the final microstate of the
system $z_B$ under the condition of energy transfer $\Delta Q$,
corresponding to the created entropy $\Delta S$, eq.(\ref{clausius}), (J2).
He states the central result of his paper:
\begin{equation}
\frac{P_+(z_B,+\Delta S|z_A)}{P_-(z_A^*,-\Delta S|z_B^*)}=e^{\Delta
S}\hspace{2.5cm} (J4)\label{jarzcentral}
\end{equation} where $P_-(z_A^*,-\Delta S|z_B^*)$ is the probability for
the time reversed process with the opposite energy transfer $-\Delta Q$
from the point in phase $\Psi$ of $\psi$ with $z_B^*=\{q_B,-p_B\}$ back to
the point $z_A^*=\{q_A,-p_A\}$ which is conjugate to the initial point
$z_A$.

Already Clausius \cite{clausius1865} used $\Delta S>\Delta Q/T=\Delta S_b$
in his formulation of eq.(\ref{clausius}) instead of the equality sign for
a ``nicht-umkehrbaren'' ($\sim$ non-reversible) process. It is quite clear,
that a fast process like the one Jarzynski considers is in general a
non-reversible fast processes  and additional to the entropy generated in
the bath there is a considerable {\em internal} entropy generation
$\Delta_iS\propto N$ in the system $\psi$ (in its phase space $\Psi$) which
is uncontrolled by any experimental realization.

Of course Jarzynski insists of heaving the system $\psi$ under absolute
control. He assumes that {\em all} its degrees of freedom $z(t)$ being
controlled at any time. For a complex system this is in general impossible.
Even for the example of Jarzynski's figure (1) he needs to control the
positions and momenta $z(t)$ of all $N=6$ particles, i.e. $36$ dofs. That
is the reason why Clausius uses in equation (\ref{clausius}) the $\ge$
sign:
\begin{equation}
\Delta S\ge \Delta S_b= \Delta
Q/T,\hspace{1.5cm}(J2_{corrected})\label{clausius1}
\end{equation} and speaks of an {\em uncompensated metamorphosis}
$\Delta_i S =\Delta S-\Delta S_b$ in \cite{clausius1854}. This is clearly
said by Prigogine when he writes for the entropy change of a complex
system:
\begin{eqnarray}
\Delta S&=&\Delta_iS + \Delta_e S \nonumber\\ \Delta_i S&\ge& 0.
\label{secondlaw}
\end{eqnarray}
Equation (\ref{secondlaw}) is the microscopic form of the second law valid
for the internal entropy generation. The external entropy change
$\Delta_eS$ can be positive, negative, or zero.

Now applied to our situation, there is no heat transfer to a reservoir
independently of whether the reservoir is small or large, i.e. $\Delta
Q\equiv 0$. The considered ensemble of trajectories (processes) in the
phase space $\Psi$ of the system $\psi$ degenerates to a single
deterministic path from $z_A$ to a single final point $z_B$, even if the
final volume has expanded. Evidently the system is not equilibrized in its
final state $B$. The backward process ($z_B^*\rightarrow z_A^*$) has to
{\em start from a non-equilibrium situation $z^*_B$}. This is in general a
{\em fractal} distribution in the systems phase space $\Psi$. However, in
any real process using a system $\psi$ with several dofs this can usually
not be done.

If, however, we allow the system to equilibrize in the new container, in
order to test the final free-energy and thus $\Delta F$ in
eq.(\ref{jarzynski}), as is explicitly assumed in Jarzynski's chapter V,
the entropy of our system $\psi$ would rise by the above $\Delta S= N
\ln(V_1/V_0)$ but still no work performed nor heat transferred in clear
contradiction to eqs.(\ref{clausius} and J2). Then equations (J51,J60) are
violated.

The uncompensated {\em internal} metamorphosis of a system $\psi$ with
several dofs leads to a surplus of internal entropy production over the
heat-transfer to the environment divided by the temperature in any
non-adiabatic process \cite{clausius1854}. This is independent of whether
there is a heat bath or not. Then in Jarzynski's identity (J51)
(\ref{jarzynski}) the left side is in general smaller than the right:
\begin{equation}
\left<\exp(-\beta W)\right>\le  \exp(-\beta \Delta
F)\hspace{0.5cm}(J51_{corrected})
\end{equation}

The misuse of Clausius' definition of entropy (\ref{clausius}) is quite
common and is further illuminated in \cite{gross214}.


\end{document}